\documentclass[11pt,twoside]{article}
\usepackage{asp2010}

\resetcounters

\begin{document}

\title{Structure and shaping processes within the extended atmospheres
of AGB stars}
\author{M. Wittkowski$^1$, D. A. Boboltz$^2$, I. Karovicova$^1$, K. Ohnaka$^3$,
A. E. Ruiz-Velasco$^{1,4}$, M. Scholz$^{5,6}$, and A. Zijlstra$^7$
\affil{$^1$ESO, Karl-Schwarzschild-Str. 2, 85748 Garching bei M\"unchen, 
Germany}
\affil{$^2$ US Naval Observatory, 3450 Massachusetts Avenue, NW, Washington, DC 20392-5420, USA}
\affil{$^3$ Max-Planck-Institut f\"ur Radioastronomie, Auf dem H\"ugel 69, 53121 Bonn, Germany}
\affil{$^4$ Departamento de Astronom{\'i}a, Universidad de Guanajuato, Apartado Postal 144, 
36000 Guanajuato, Mexico}
\affil{$^5$ Zentrum f\"ur Astronomie der Universit\"at Heidelberg (ZAH), Institut f\"ur Theoretische 
Astrophysik, Albert-Ueberle-Str. 2, 69120 Heidelberg, Germany}
\affil{$^6$ Sydney Institute for Astronomy, School of Physics, University
of Sydney, Sydney, NSW 2006, Australia}
\affil{$^7$ Jodrell Bank Centre for Astrophysics, School of Physics 
and Astronomy, University of Manchester, Manchester M13 9PL, UK}
}

\begin{abstract}
We present recent studies using the near-infrared instrument AMBER of the
VLT Interferometer (VLTI) to investigate the structure and shaping processes 
within the extended atmosphere of AGB stars.
Spectrally resolved near-infrared AMBER observations of the Mira variable
S~Ori have revealed wavelength-dependent apparent angular sizes.
These data were successfully compared to dynamic model atmospheres, 
which predict wavelength-dependent radii because of geometrically
extended molecular layers.
Most recently, AMBER closure phase measurements of several AGB stars
have also revealed wavelength-dependent deviations from 0/180 deg., 
indicating deviations from point symmetry. The variation of closure phase
with wavelength indicates a complex non-spherical stratification 
of the extended atmosphere, and may reveal whether observed asymmetries 
are located near the photosphere or in the outer molecular layers.
Concurrent observations of SiO masers located within the extended
molecular layers provide us with additional information on the 
morphology, conditions, and kinematics of this shell.
These observations promise to provide us with new important insights into 
the shaping processes at work during the AGB phase. With improved imaging 
capabilities at the VLTI, we expect to extend the successful story of 
imaging studies of planetary nebulae to the photosphere and extended 
outer atmosphere of AGB stars.
\end{abstract}

\section{Introduction}
Asymptotic Giant Branch (AGB) stars are low-
and intermediate mass stars, such as our Sun, in their final phase
of evolution that is driven by nuclear burning. Mass-loss becomes
increasingly important toward the tip of the AGB evolution, when
the ``superwind'' phase occurs. The superwind mass-loss reduces
the convective stellar envelope until the star starts to shrink.
Then the star evolves at almost constant
luminosity to higher effective temperatures, passes the post-AGB
phase, and becomes a planetary nebula (PN). In this phase the now hot 
inner star ionizes its envelope, which is a remnant of the superwind 
mass-loss during its AGB and post-AGB phases (e.g. Habing \& Olofsson 
\citealt{habing03}).

While the mass-loss process during the AGB phase is the most important
driver for the further stellar evolution toward the PN phase, the details
of the mass-loss process and its connection to the structure of the
extended atmospheres and the stellar pulsation are not well understood and are
currently a matter of debate, in particular for oxygen-rich AGB stars
(e.g. Woitke et al. \citealt{woitke06}, H\"ofner \& Andersen \citealt{hoefner07}).

In the past decade, imaging studies of
PNe and protoplanetary nebulae (pPNe) have revealed a great
diversity of morphologies. This variety of shapes of pPNe and PNe
is seemingly caused by processes at the end of the
AGB evolution, but the details of the shaping processes
are not well understood (e.g. Balick \& Frank \citealt{balick02}).

Interferometric techniques at optical and radio wavelengths
have proven their ability to provide important observational constraints 
on the atmosphere and mass-loss process of AGB stars by resolving the stellar disk
and the circumstellar environment (e.g. Quirrenbach et al. \citealt{quirrenbach92},
Reid \& Menten \citealt{reid97}, Kemball \& Diamond \citealt{kemball97},
Perrin et al. \citealt{perrin04}). Deviations from circular symmetry have
been detected in the circumstellar environment (CSE) of AGB stars at
radio as well as optical wavelengths (e.g. Reid \& Menten \citealt{reid07},
Ragland et al. \citealt{ragland08}).

Most recent studies using the near-infrared (AMBER) and mid-infrared (MIDI)
instruments of the VLT Interferometer (VLTI) have added 
important information to our understanding of the pulsation and mass-loss
of AGB stars thanks to their spectro-interferometric capabilities
(Ohnaka et al. \citealt{ohnaka05,ohnaka06,ohnaka07}; 
Wittkowski et al. \citealt{wittkowski07,wittkowski08}; Le Bouquin et al. \citealt{lebouquin09};
Chiavassa et al. \citealt{chiavassa10}).

Here, we describe our most recent observations of the structure and 
morphology of the extended atmosphere
of AGB stars using the VLTI/AMBER instrument, and their implication
on the mass-loss process and the onset of asymmetric shapes during the
AGB evolution.

\section{Observations}
\label{sec:observations}
The first AMBER observations of an AGB star were obtained on 12 October 2007
on the Mira variable S~Ori (period 414 days, Samus et al. \citealt{samus09}), 
and are 
described in Wittkowski et al. (\citealt{wittkowski08}).
These observations used the low resolution mode of AMBER with a spectral resolution
of $\sim$35 and utilized the fringe tracker FINITO and three VLTI Auxiliary Telescopes 
(ATs) positioned on stations E0, G0, and H0 with ground baselines of 16\,m, 32\,m, 
and 48\,m.
Since then, we have obtained additional AMBER observations of the Mira variables
S~Col (325 d), T~Col (225 d), W~Vel (394 d), RW~Vel (443 d), R~Cnc (361 d), 
X~Hya (301 d), and RR~Aql (394 d) between September 2008
and June 2010. These observations used in addition to the low resolution mode
the medium resolution modes of 
AMBER with a spectral resolution of $\sim$1500 in the near-infrared $H$ and $K$
bands, and also utilized additional AT configurations. These observations will
be described in detail in forthcoming papers. Some of these observations were
coordinated with concurrent VLBA observations of the SiO and H$_2$O maser emission.
In addition to Mira variables, AMBER data on OH/IR stars have been obtained
in April 2008 (Ruiz Velasco et al., these proceedings). 
\section{Modeling}
Few dynamic atmosphere models for oxygen-rich Mira stars
are available. The P and M model series (Ireland et al. \citealt{ireland04a,ireland04b})
are complete self-excited dynamic
model atmospheres of Mira stars designed to match the prototype
oxygen-rich Mira stars $o$~Cet and R~Leo. They have been used successfully
compared to VLTI/VINCI broadband interferometric data of $o$~Cet and R~Leo
(Woodruff et al. \citealt{woodruff04}; Fedele et al. \citealt{fedele05}).
Compared to $o$~Cet and R~Leo, our target stars have slightly different
periods, masses, and radii. However, the general model results are not expected 
to be dramatically different for our target stars compared to 
$o$~Cet and R~Leo (cf. the discussion in Wittkowski et al. \citealt{wittkowski07}).
As a result, the P and M model series were chosen as the currently best available
option to describe Mira star atmospheres. Wittkowski et al. (\citealt{wittkowski07})
have added an ad-hoc radiative transfer model to these model series to describe
the dust shell as observed by the mid-infrared interferometric instrument MIDI.
However, at near-infrared wavelengths, the contribution of the dust shell can be
neglected for our target stars. Gray et al. (\citealt{gray09}) have combined
these hydrodynamic atmosphere models with a maser propagation code in order
to describe the SiO maser observations that have been coordinated for some of 
our target stars.

\section{Results}
\begin{figure}
\centering
\resizebox{0.32\hsize}{!}{\includegraphics{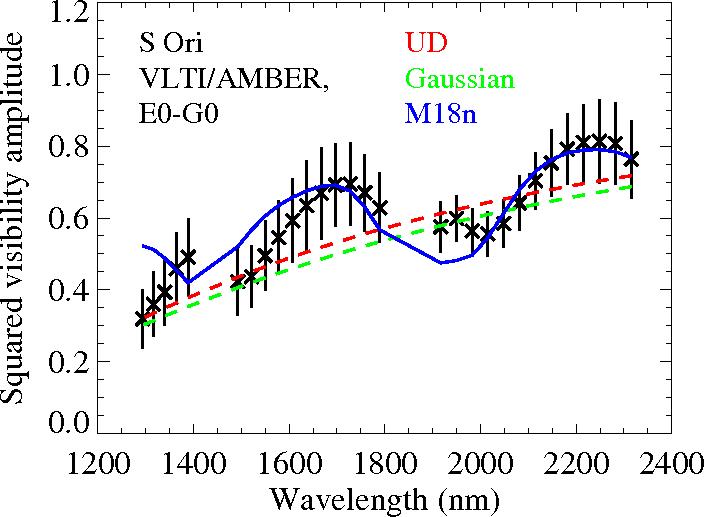}}
\resizebox{0.32\hsize}{!}{\includegraphics{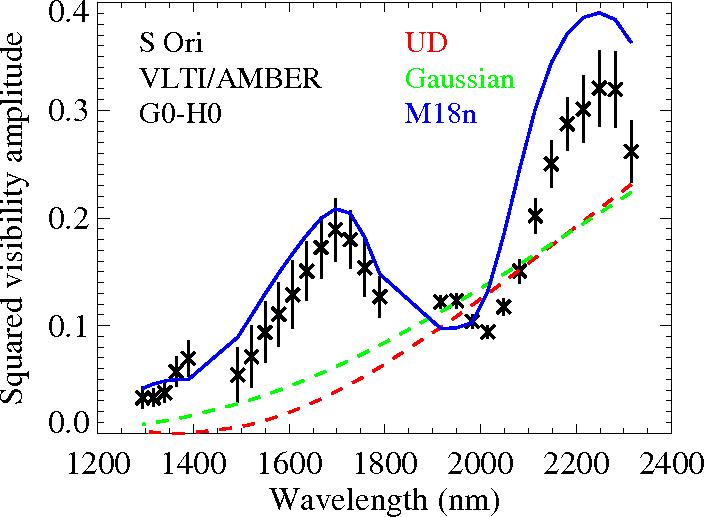}}
\resizebox{0.32\hsize}{!}{\includegraphics{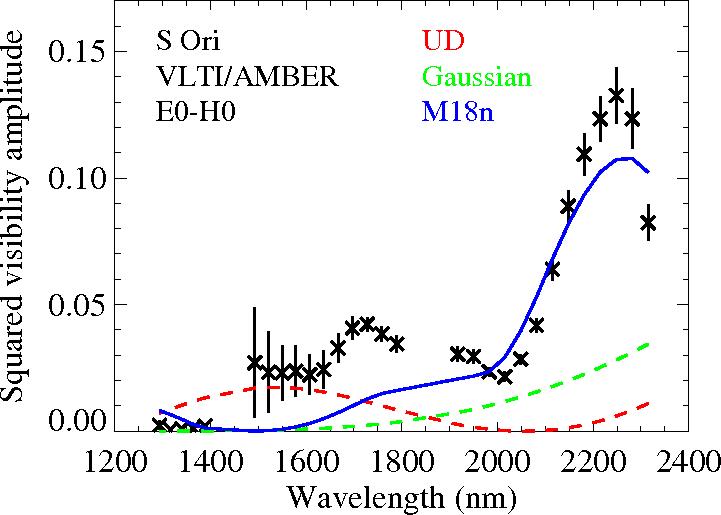}}
\caption{VLTI/AMBER visibility function of S~Ori compared to simple
models of a uniform disk and a Gaussian disk, as well as compared to the
dynamic atmosphere model. From Wittkowski et al. (\citealt{wittkowski08}).}
\label{fig:sori}
\end{figure}
\begin{figure}
\centering
\resizebox{0.32\hsize}{!}{\includegraphics{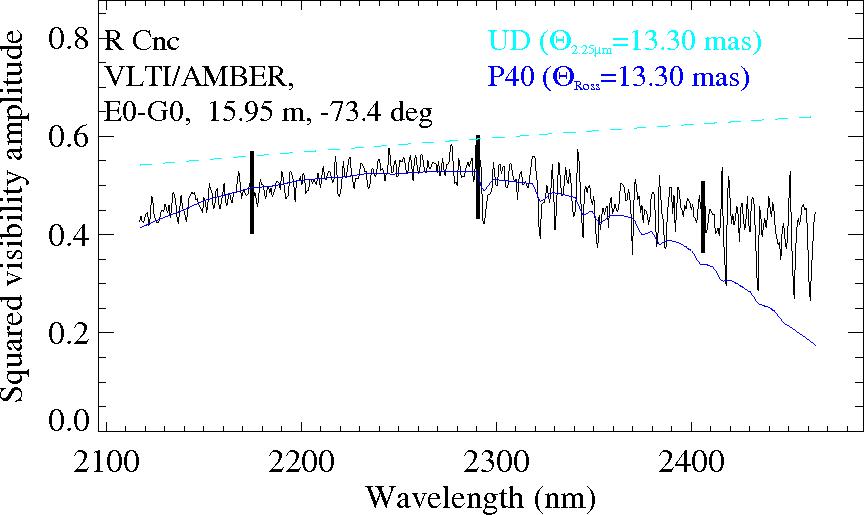}}
\resizebox{0.32\hsize}{!}{\includegraphics{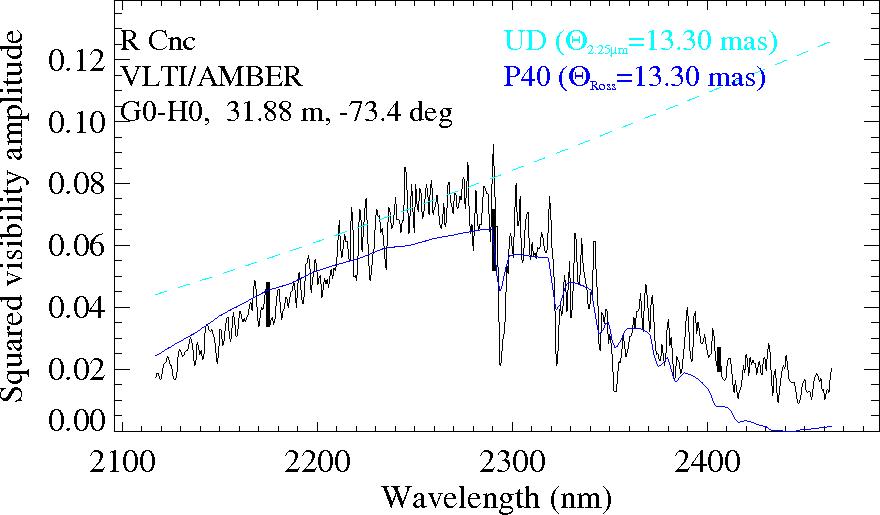}}
\resizebox{0.32\hsize}{!}{\includegraphics{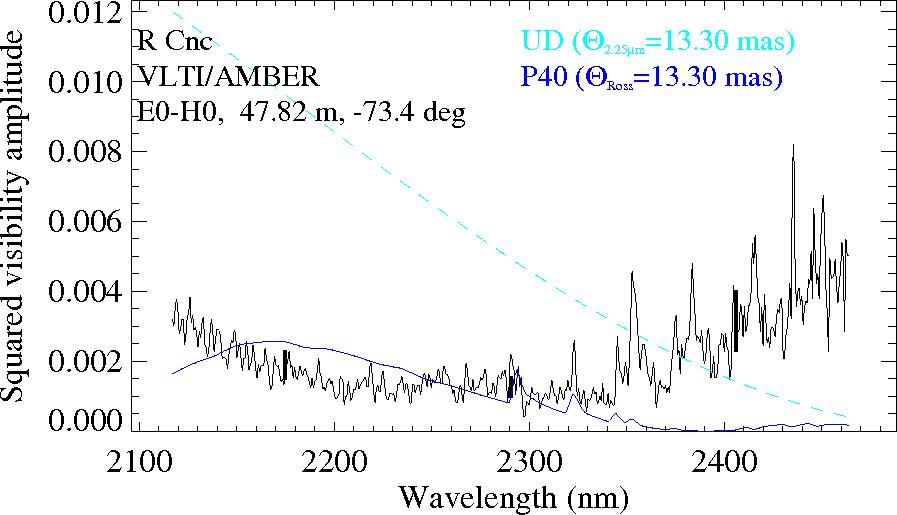}}
\caption{VLTI/AMBER visibility function of R~Cnc compared to a simple
model of a uniform disk, as well as compared to a 
dynamic atmosphere model.}
\label{fig:rcnc}
\end{figure}
Fig.~\ref{fig:sori} shows the AMBER visibility data of S Ori from
Wittkowski et al. (\citealt{wittkowski08}). 
The visibility data of S~Ori show significant wavelength-dependent
features clearly deviating from uniform disk (UD) and Gaussian models 
of constant diameter
on all three baselines. This indicates variations in the apparent
angular diameter. For comparison, $\gamma$~Eri, a regular M giant,
was observed during the same night with the same instrument settings,
and did not show such deviations from a UD curve, confirming that the
features seen for S Ori are not instrumental or atmospheric effects.

Model M18n provided the best formal fit to our S~Ori visibility
data out of the available phase and cycle combinations of the M series.
The synthetic visibility values based on the M18n model
compared to our AMBER observation
are also indicated in Fig.~\ref{fig:sori}, showing that our AMBER visibility
data could be described well by the dynamic atmosphere model series.
The deviations of the model visibilities from a uniform disk of constant diameter
is caused by molecular layers (most importantly CO and H$_2$O) lying above
the continuum-forming layers. At spectral channels, where the molecular opacity
is low, we see a larger contribution from the photosphere, and the target appears smaller. 
At spectral channels where the molecular opacity is larger, we see a larger contribution
from the extended atmospheric molecular layers, and the target appears larger.

In summary, our AMBER observations of S Ori generally confirmed
the predictions by the M model series and we found that the observed variation of
diameter with wavelength can be understood as the effect of phase-dependent water vapor
and CO layers lying above the photosphere. We also concluded that more such observations 
on more targets and at more phases were needed to confirm and constrain the model predictions 
in more detail.

Since the work described in Wittkowski et al. (\citealt{wittkowski08}), we have
obtained additional AMBER data of more targets and at more phases, as described
in Sect.~\ref{sec:observations}. Fig.~\ref{fig:rcnc} shows as an example the visibility 
function of the Mira variable R~Cnc obtained with the medium resolution model of the 
AMBER instrument (spectral resolution $\sim$1500). These data confirm the conclusion
from Wittkowski et al. (\citealt{wittkowski08}) that Mira variables show 
wavelength-dependent angular diameters when observed with spectro-interferometric
techniques that can be explained by molecular layers lying above the continuum-
forming photosphere, and that are consistent with predictions by dynamic model
atmospheres. In addition to these AMBER observations, Chiavassa et al. (\citealt{chiavassa10})
have obtained AMBER data of the very cool late-type star VX~Sgr, and Ruiz Velasco
et al. (these proceedings) have obtained AMBER data of three highly evolved
AGB stars of OH/IR type. All these data show similar characteristics of the 
AMBER visibility function. The characteristic 'bumpy' AMBER visibility curves indicating 
the presence of molecular layers, thus seems to be a common feature of evolved 
oxygen-rich stars.
\section{Shaping processes}
\begin{figure}
\centering
\resizebox{0.7\hsize}{!}{\includegraphics{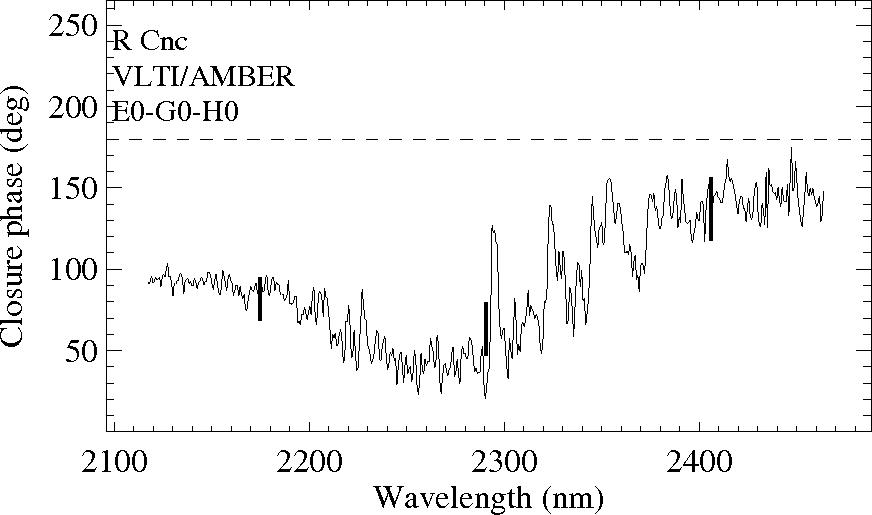}}
\caption{VLTI/AMBER closure phase of R~Cnc.}
\label{fig:rcncphase}
\end{figure}
Fig.~\ref{fig:rcncphase} shows the AMBER closure phase measurement corresponding
to the visibility curves of Fig.~\ref{fig:rcnc}. The interferometric closure phase
is a measure of the point symmetry of a source. Values of 0\deg\ and 180\deg\
indicate point symmetry, other values deviations from point symmetry.
R~Cnc shows closure phase values that are significantly different from
0\deg and 180\deg, thus indicate a significant deviation from point symmetry.
The R~Cnc closure phases vary with wavelength. At wavelengths around 2.25$\mu$m,
where the visibility values show a maximum corresponding to a small angular 
diameter, the closure phase shows values of about 40\deg. Shortward of
2.25$\mu$m, where the visibility decreases and where the water vapor 
opacity becomes larger, the closure phase increases to 90\deg. 
Longward of 2.25$\mu$m, where the visibility decreases both due to 
CO and water vapor, the closure phase increases to 150\deg, with closure
phase peaks marking the peaks of the CO bandheads. A similar characteristic
signal of the closure phases are seen for other targets of our sample.

The interpretation of the closure phase measurements is work in progress.
They might indicate a complex non-spherical stratification 
of the photosphere and extended atmosphere, and may reveal whether observed asymmetries 
are located near the photosphere or in the outer molecular layers.
These observations thus promise to lead to new insights into 
the shaping processes at work during the AGB phase.
As an example, the detection of photospheric convection cells
and corresponding clumps in the molecular layer that have characteristics
as those predicted by 3D atmosphere models (e.g. Freytag \& H\"ofner \citealt{freytag08})
would point to a process of large-scale photospheric convection.
Random clumps in the molecular shell
may point to highly temporally and also spatially variable chaotic mass ejections
caused by pertubations in the oscillations (Icke et al. \citealt{icke92}) or
by magnetic fields (e.g. Suzuki \citealt{suzuki07}).

\section{Summary}
AMBER spectro-interferometry shows wavelength-dependent apparent sizes of 
AGB stars, including Mira variables and OH/IR stars. These observations are
generally consistent with dynamic model atmosphere predictions that include
molecular layers, in the infrared most importantly CO and H$_2$O, lying above
the continuum-forming photosphere. AMBER closure phase measurements of a sample
of resolved Mira variables indicate deviations from point symmetry. 
A characteristic wavelength dependence of the closure phase values indicates
a complex non-spherical stratification of the extended atmosphere.
A detailed interpretation of these measurements is needed, and may reveal
whether asymmetric structures originate at the photosphere or at more extended
layers, and how they develop from smaller to larger scales. Advanced imaging
capabilities with the VLTI thanks to improved baseline configurations and
2nd generation instruments combining a larger number of beams will allow us
to obtain model independent images of the extended atmospheres of AGB stars.
We are thus at a turning point in the history of knowledge of AGB stars,
where imaging studies can be extended from pPNe and PNe
to layers of AGB stars at the photosphere and at molecular layers
close to the photosphere, and thus reveal the shaping processes at
work during the AGB phase.



\end{document}